\documentclass[aps,prl,twocolumn,groupedaddress,amsmath,amssymb]{revtex4-1}
\usepackage{graphicx}  % needed for figures
\usepackage{dcolumn}   % needed for some tables
\usepackage{bm}        % for math
\usepackage{verbatim}  % for math

\newtheorem{myDefinition}{Definition}
\newtheorem{myExample}{Example}

\begin{document}

\title{Defining the Observed World in Quantum Mechanics}
\author{Hitoshi Inamori}
\email{Previous Academic affiliation: Centre for Quantum Computation, Clarendon Laboratory, Oxford University}
\affiliation{
   Societe Generale Securities, Tokyo Branch\\
   Palace Building, 1-1-1 Marunouchi, Chiyoda-ku, Tokyo 100-8206 Japan
   }
\date{\today}
\begin{abstract}
This paper defines what constitutes the Observed World in the Quantum Mechanical framework, based strictly on what is actually observed beyond doubt, instead of building observables on what is inferred from actual observations. Such principle narrows down considerably what can be considered as being part of the Observed World. On the other hand, we argue that some information -- that is in general assumed as granted -- should actually be considered as being part of the Observed World. We discuss the implications of such assertion, in the way we perceive time evolution, information growth and causality.  
\end{abstract}
\maketitle

\section{Introduction}
In any physical experiment, the experiment outcomes are generally described by observable quantities which are introduced ad-hoc, given the experimental setup and the physical system under study. If a measuring apparatus is part of an experimental setup, then, based on the assumed functioning of that apparatus, we introduce an associated observable quantity for the system under experiment.
For instance, in an experiment involving a photo-detector, we introduce the observable ``the presence of a photon at the photo-detector''. Indeed, based on our past experience and our interpretation of physics, we assume that we hear a click each time one or several of so called ``photons'' reach the detector. Therefore, we infer the observable ``presence of photon at the photo-detector'' in an experimental setup involving such device.

There are many assumptions in asserting that this observable ``presence of photon at the photo-detector'', is correctly describing the reality: not only the photo-detection is subject to noise and appartus failure, but the very notion of photon, and the functioning of the photo-detection apparatus, are based on our current understanding of physics. More prosaically, it is not impossible that the photo-detector has been altered maliciously without our knowledge and that the detector clicks at some arbitrary time interval. Therefore, rigorously speaking, when the detector clicks, the only thing which we are sure of, is that a click is heard. 

The stance in this paper is to rule out any inferred observed quantities (such as ``presence of photons'' when a click is heard), and to define the Observed World, as being the set of all quantities which are actually observed (like the hearing of the click), which are beyond any doubt and which are not subject to any interpretation. 
Given such restriction, the Observed World is much smaller than what we usually associate with the notion of a ``world'', and we present few consequences of this principle.

\section{What is the Observed World ?}
We start with the definition of the Observed World and describe what it means in practice. 

\begin{myDefinition}
The Observed World is the set of all quantities, that are actually observed. 
\end{myDefinition}

By ``actually observed'', we exclude any results which are inferred, or implied from actual observations. We only consider observations which arise directly into awareness, i.e. without any possibility of doubt, and which is not subject to any interpretation or assumption. 

As such, reported observation, by a measuring apparatus, or by another experimenter, is not part of the Observed World : the only observable in this case is the signal (sound or visual signal for instance), coming from the measuring apparatus or the other experimenter. More precisely, the observable quantity is the physical input that comes into awareness. The assumed causes, leading to this physical input into awareness, are not part of the Observed World, as they are not the inputs which are directly available to the awareness. Likewise, in the example of the introduction, the ``presence of photon at the photo-detector'' is not part of the Observed World. The only related quantity in the Observed World is the click which is heard.

On the other hand, all information which come into awareness are part of the Observed World. As such, memory content (in its broad sense), which comes into awareness, is, by definition, part of the Observed World : memory output is an observed physical signal, on an equal footing with any signal that comes from sensory organs such as eyes or ears.  
Research in neuroscience~\cite{Libet, Soon} suggests that -- at least in some situations -- decision making is completed in the brain, before the result of such decision making is known to awareness. If such mechanism holds in general, then decision making outcome is also part of the Observed World. The notion of free will is replaced by the observation of a physical process, in which a component of the brain -- which is outside awareness -- computes the decision. Not unlike Schr\"odinger's cat, that decision is only materialized once it is observed by awareness. 

The exact set of observable quantities available to awareness remains debatable as the physical frontier of consciousness is still under investigation. However, in practice, we assume that the Observed World is the set of all the inputs which are available to the part of the brain which is responsible for consciousness, provided that such part of the brain can be identified and delimited in practice.

Judging from the discussion above, the Hilbert space, describing the Observed World in the quantum mechanical framework, has limited dimensionality: surely the part of the brain responsible for consciousness is limited in size, and so is the dimension of the Hilbert space associated with it. 
On the other hand, the Observed World is, by definition, the set of all the actually observable quantities. Whether the universe is assumed to be much bigger than the Observed World or not, the Observed World gives exactly all the information that are actually observable from the universe. 

\section{On the size of the Observed World}
In the previous section, we stated that the Observed World represents actually everything which can be known, and is known, from the universe, and that the dimension of the Hilbert space associated with the Observed World is limited in size. Any observation may find the state of the Observed World in one of the basis state, but the choice of the state in this basis is limited by the size of the basis. 
In the framework of the ``relative state'' formulation of quantum mechanics~\cite{Everett}, one often argues that such theory engenders a never-ending split of the universe into infinite branches of so-called ``multiverses''~\cite{DeWitt}. In our view, the ``relative state'' theory does not lead to such an inflation of the universe into ``multiverses''. Because the Observed World is limited in size, the amount of information it contains is also limited. One extra information available to awareness is added in detriment of another information which is lost. As an old adage goes, ``learning is forgetting''.

The important thing to note here is that memory is also part of the Observed World. The usual description of experiments in physics (quantum physics included), considers that the results of past observations are intrinsic truths: if an experiment yields a certain result, say A, then we usually consider the result of such experiment as granted after the experiment. As natural as this description may seem, it is based on an implicit assumption that such past results are stored in some memory which is available to awareness. 
We could argue that the result A could be stored in some physical system which can act as memory outside awareness. However, our view is that such memory, as long as it is does not appear to awareness, can be altered without being noticed. Therefore, we cannot take the result A as granted, as no observation can guarantee that A is true. 
Our point is that implicit assumption on past observations should be ruled out and described explicitly. Past results which are available to awareness must be described as part of the Observed World. The part of the Observed World corresponding to past results is called `` memory''. As any observed quantity which is part of the Observed World, memory's output is part of the description provided by the state vector associated with the Observed World : a correct modelisation should describe memory as a quantum mechanical system, like any other physical system that is observed.

\begin{myExample} Take a very simple fictive example in which there are two die 1 and 2 and that the Observed World is represented by a 12-dimensional Hilbert space. Obviously, the Observed World cannot encode the value returned by both die (which would require 6 $\times$ 6 = 36 dimensions). For instance, the Observed World can be observed in one of the 12 states: Dice 1 is observed and its value is 1, 2, 3, 4, 5 or 6. Or Dice 2 is observed and its value is 1, 2, 3, 4, 5 or 6. Whatever the measurement applied to the Observed World, one cannot get the value of both die. Endeavoring to learn about the value returned by Dice 2 will necessarily degrade the information we get about Dice 1. One could argue that Dice 1 can be observed, the result of such observation is stored in the memory, and then Dice 2 is observed. Our point is that in such a case, one needs to expand the Observed World to include the state of the memory. Memory will itself be in one of the 12 states (encoding the dice chosen and value observed) and the Observed World’s dimension will grow to 12$\times$12 = 144, larger than the minimum 36 needed to encode both die.
\end{myExample}

\section{On the perception of time}
Memory is part of the Observed World. As any quantum mechanical system, it is described within a Hilbert space. In the case of memory, each basis state of the associated Hilbert space corresponds to the remembrance of a different history, of a different past. And each possible outcome coming from the observation of the memory corresponds to a possible realized history. 

\begin{myExample} To illustrate this idea, assume a world in which one can only observe the values returned by a set of die. No dice is observed at the beginning. The value returned by the first dice is observed at time 1, then the value returned by the second dice is observed at time 2, etc. 
This can be described by expanding the previous example to $n$ die. Assume that the Observed World tells how many die have been observed starting with Dice 1 then continuing with the next Dice 2, etc. Suppose that the Observed World encodes also the values observed for each dice that has been looked at. 

The dimension of the Hilbert space describing the Observed World is $\sum_{k=0}^n 6^k$
(to encode the value of each dice, given that $k$ die are observed, for $k=0, 1, 2, \ldots , n$). Measurement of the Observed World can theoretically lead to any state among the $\sum_{k=0}^n 6^k$ states. Depending on the outcome of this measurement, not only the values of the observed die will differ, but also the very fact that a given dice has been observed or not. Said differently, it is the outcome of this measurement which tells what instant of time is being observed. 
\end{myExample}

Coming back to our main discussion, anything which is beyond what is actually observed from the memory, i.e. from the Observed World, is irrelevant: as Russell noted~\cite{Russell}, ``there is no logical impossibility in the hypothesis that the world sprang into being five minutes ago, exactly as it then was, with a population that ``remembered'' a wholly unreal past.'' Memory content only tells what the memory content is, and nothing beyond.

Now, memory (again, taken in its broad sense) is central in establishing what we call ``causality''. Most, if not all, of the causality rules which arise to consciousness, including the laws of physics, are based on the outputs from memory. Depending on the memory outputs, causality laws may potentially take very different forms.

Including memory in the Observed World impacts also the way we explain time perception. Conscious perception of past and present depends on the outputs resulting from the observation of memory: ``past'' is the set of events which are reported by the memory as having happened. Likely, the present is constituted by events which are reported by the memory as having just happened. Awareness seems to link memory states which are historically coherent, in the sense that a memory state in which an event is about to occur appears ``to be followed'' by another memory state in which such event is occurring. However, this paper's view is that such a flow does not actually exist: any eligible memory state is a potential reality, without any fundamental ``ordering'' between them. Time perception, as any causality rule, is a byproduct of the observation of the memory. And memory is part of the Observed World.

\begin{myExample} In the previous example, a measurement of the Observed World can lead to the state in which 3 die have been observed with the outcome 1 for the 1st dice, 4 for the 2nd, and 5 for the 3rd. The measurement could also have lead to the state in which 2 dices have been observed with the outcomes 1 for the 1st dice, 4 for the 2nd dice. The second outcome seems to be interpreted as a coherent ``past'' to the first outcome. However, there is no fundamental relationship between the two outcomes. Perception of time flow or causality linking the two measurement outcomes is not based on a fundamental ``dynamics'' of the universe, but is rather a result of interpretation or perception.
\end{myExample}

\section{Conclusion}
In this paper, our main objective is to identify what is indisputably observed in any physical experiment. We conclude that what is ultimately observed is the set of all the signals which come to awareness. We defined the ``Observed World'' as being the set of such signals. 
 
We assert that what we call ``past'', ``time flow'' and ``causality'' are the byproducts of a subset of the Observed World. Such subset is identified as being what we commonly call ``memory''. The view presented in this paper is that our perception of time, causality and physical laws, are dictated by the output of that part of the Observed World.

\end{document}